\documentclass[12pt]{article}

\usepackage{graphicx}
\usepackage{amsmath}
\usepackage{textgreek}
\usepackage{times}
\usepackage{color}
\usepackage{physics}
\usepackage{dirtytalk}

\usepackage{times}

\title{Large-scale time-multiplexed nanophotonic parametric oscillators} 

\author
{Robert M. Gray$^{1\ast}$, Ryoto Sekine$^{1\ast}$, Luis Ledezma$^{1}$, Gordon H. Y. Li$^{2}$, Selina Zhou$^{1}$, \\ Arkadev Roy$^{1}$, Midya Parto$^{1}$, and Alireza Marandi$^{1\dagger}$\\
\\
\normalsize{$^{1}$Department of Electrical Engineering, California Institute of Technology,}\\
\normalsize{Pasadena, CA 91125, USA}\\
\normalsize{$^{2}$Department of Applied Physics, California Institute of Technology,}\\
\normalsize{Pasadena, CA 91125, USA}\\
\\
\normalsize{$^\ast$ These authors contributed equally to this work.}\\
\normalsize{$^\dagger$Email: marandi@caltech.edu}
}

\date{}

\begin{document} 


\baselineskip24pt


\maketitle 


\begin{abstract}
Arrays of nonlinear resonators offer a fertile ground for a wide range of complex phenomena and opportunities for advanced photonic sensing and computing. Recently, significant attention has focused on studying coupled resonators in special-purpose configurations either on chips or in table-top experiments. However, a path to realizing a large-scale programmable network of nonlinear photonic resonators remains elusive because of the challenges associated with simultaneously achieving strong nonlinearity, independent operation of the resonators, and programmability of the couplings. In this work, we break these barriers by realizing large-scale, time-multiplexed optical parametric oscillators (OPOs) on a single lithium niobate nanophotonic chip. We show independent operation of 70 identical OPOs in an ultrafast nanophotonic circuit. The OPOs exhibit an ultra-low threshold of a few picojoules, substantially surpassing the strength of nonlinearity of other platforms. Using our ultrafast nanophotonic circuit, a network of N OPOs with programmable all-to-all couplings requires only a few additional components. The time-multiplexed nanophotonic OPOs can enable myriad applications, including ultrafast classical and quantum information processing.
\end{abstract}

\section*{Introduction}

Nonlinear resonators are emerging as one of the most versatile building blocks for a wide range of photonic systems benefiting applications in quantum information processing \cite{asavanant2019generation, larsen2019deterministic}, stochastic computing \cite{marandi2012all, roques2023biasing}, metrology \cite{obrzud2019microphotonic, spencer2018optical}, and spectroscopy and sensing \cite{zhang2013mid, stern2020direct}, among others. Coupled nonlinear resonators further promise broad potentials, which have been showcased through a variety of table-top experiments, for instance using optical parametric oscillators (OPOs) \cite{mcmahon2016fully, takeda2017boltzmann, honjo2021100, roy2023non} and lasers \cite{arwas2022anyonic, leefmans2022topologicalmode}. However, in nanophotonics, demonstrations of coupled nonlinear resonators remain in the small-scale regime \cite{pal2024linear, zhang2019electronically, jang2018synchronization, okawachi2020demonstration} or suffer from limited programmability \cite{bandres2018topological, mittal2021topological}.

OPOs using quadratic nonlinearity are one of the most promising nonlinear photonic resonators, with a long history as table-top tunable sources in hard-to-access wavelength ranges \cite{dunn1999parametric}. More recently, OPOs have been used for a wide range of applications spanning from frequency comb spectroscopy \cite{muraviev2018massively} to sensing \cite{gray2023cavity}, quantum information processing \cite{konno2024logical}, and computing \cite{wang2013coherent, marandi2014network, yamamoto2017coherent}.  Advances in thin-film lithium niobate have enabled realization of nanophotonic OPOs \cite{mckenna2022ultra,lu2021ultralow,ledezma2022intense} with substantial miniaturization and threshold enhancement due to the sub-micron modal confinement. Synchronous pumping of dispersion-engineered nanophotonic OPOs with ultrashort pulses is particularly important because it leads to ultralow-threshold operation \cite{roy2023visible} enabling opportunities for energy-efficient ultrabroad comb sources \cite{sekine2023multi} and quantum information processing \cite{yanagimoto2023mesoscopic}.

Ultrafast nanophotonic OPOs not only benefit from ultra-low-energy operation, they also enable large-scale time-domain multiplexing (TDM) for realization of programmable OPO networks. Time-multiplexed resonator networks have been demonstrated on table-top experiments for a wide range of studies in optical computing \cite{li2023photonic, marandi2014network, honjo2021100}, topological photonics \cite{leefmans2022topological, parto2023non}, and non-equilibrium phase transitions \cite{roy2023non}, among others. Compared with other multiplexing schemes for realizing nonlinear resonators, such as spatial and spectral multiplexing \cite{ozawa2019topological}, TDM benefits from scalability, and the strength of nonlinearity \cite{leefmans2022topologicalmode}.

Here, we demonstrate the first nanophotonic realization of large-scale time-multiplexed OPOs in lithium niobate (LN). By leveraging a large parametric gain and dispersion engineering, we achieve simultaneous oscillation of as many as 70 independent OPOs at a 17.5 GHz repetition rate, limited primarily by the speeds of our pump repetition rate and detection electronics. We verify the independence of the oscillators through an interferometric measurement at the output of the chip which confirms the vacuum-seeded randomness of each oscillator.

\section*{Results}

The on-chip time-multiplexed OPO system is schematically depicted in Fig. 1A. The temporally separated resonators, equivalent to the N independent resonators shown in Fig. 1B, share the same optical path length and long periodically poled section. High gain provided by the periodically poled region ensures a low on-chip threshold pulse energy of a few pJ for the ps pulses used in the experiment. Adiabatic tapers couple more than 96\% of the signal light into the resonator while ensuring very little coupling (\(\leq\) 2\%) for the pump. Finally, dispersion engineering of the waveguide geometry ensures near-zero group velocity mismatch between the pump and signal during the nonlinear interaction to achieve a high gain and gain bandwidth as well as near-zero group velocity dispersion for the signal in the roundtrip, which preserves the short-pulse operation necessary for maintaining the independence of the oscillators.

We operate in both the non-degenerate (Fig. 1C) and degenerate (Fig. 1D) regimes by adjusting the pump frequency detuning with respect to the cavity. In the non-degenerate regime, pump photons are split into signal and idler photons at different frequencies, and their phase relationship is given by:

\begin{equation}
    \phi_p - \phi_s -\phi_i = \frac{\pi}{2},
\end{equation}

\noindent where \(\phi_{p}\), \(\phi_{s}\), and \(\phi_{i}\) are the pump, signal, and idler phase, respectively. Taking the pump phase as determined, this relationship leaves the signal and idler phases free, constraining only their sum. Thus, as illustrated in the phase space diagram of Fig. 1C, wherein the radial coordinate represents the pulse amplitude and the angle represents the phase, the phases are random and may take on any value.

\begin{figure}[hp!]
\centering\includegraphics[width=16cm]{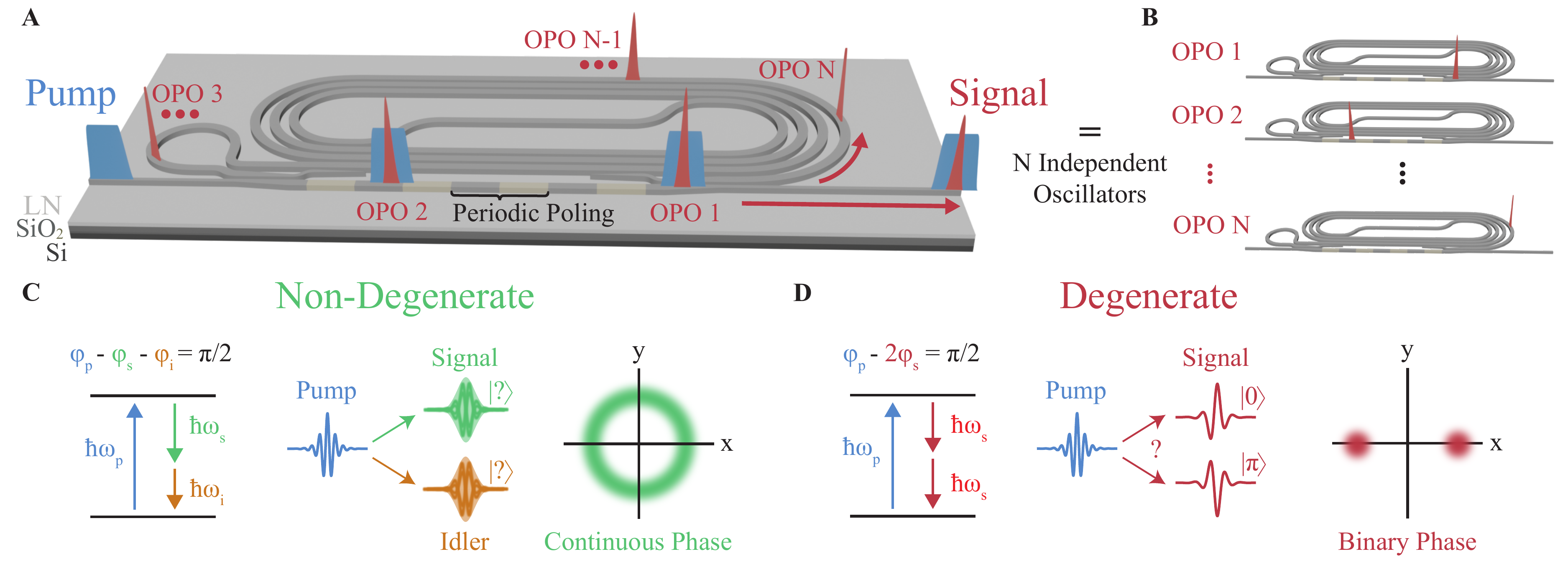}
\caption{{\bf Time-multiplexed nanophotonic OPOs.} ({\bf A}) Schematic of the device based on thin-film lithium niobate. N pulses in a long cavity are equivalent to ({\bf B}) N independent time-multiplexed oscillators. ({\bf C}) In the non-degenerate regime, the pump photons split into signal and idler photons at different frequencies. In this case, the phase of each is unconstrained, as shown in the signal phase space diagram. ({\bf D}) In the degenerate regime, the pump photons split into indistinguishable signal photons at the half-harmonic of the pump, resulting in a binary phase for the signal.}
\end{figure}

By contrast, in the degenerate regime, the signal and idler both resonate at the half-harmonic of the pump and thus are indistinguishable from one another. In this case, equation 1 consolidates to: 

\begin{equation}
    \phi_p - 2\phi_s = \frac{\pi}{2},
\end{equation}

\noindent such that signal phase is restricted to one of two phase states (Fig. 1D), which we refer to as \(\ket{0}\) and \(\ket{\pi}\).

Figure 2 shows the measurement protocol for ensuring independence of the N oscillators in the time-multiplexed OPO cavity. The measurement setup is shown in Fig. 2A and is described in further detail in Supplementary Information Section 1. The cavity is pumped by ps pulses at 1045 nm generated by an electro-optic (EO) comb with a variable repetition rate. TDM is achieved through selecting a pump repetition period, T\textsubscript{rep}, that is a harmonic of the cavity roundtrip time, T\textsubscript{RT}, meaning T\textsubscript{rep} = T\textsubscript{RT}/N, such that N OPOs are made to oscillate simultaneously in the long spiral cavity. In our system, the 53-cm spiral results in a round-trip time of 4 ns and corresponding 250-MHz FSR. The signal is coupled into an unbalanced Mach-Zehnder interferometer (MZI), the output of which is collected on a photodetector, which allows for measurement of the relative phases of the output pulse train. A 92:8 splitter placed before the MZI additionally sends a fraction of the signal to a reference photodetector for normalizing the measurement. The inset shows a microscope image (left) and SEM image (right) of a portion of the long spiral resonator.

\begin{figure}[hp!]
\centering\includegraphics[width=16cm]{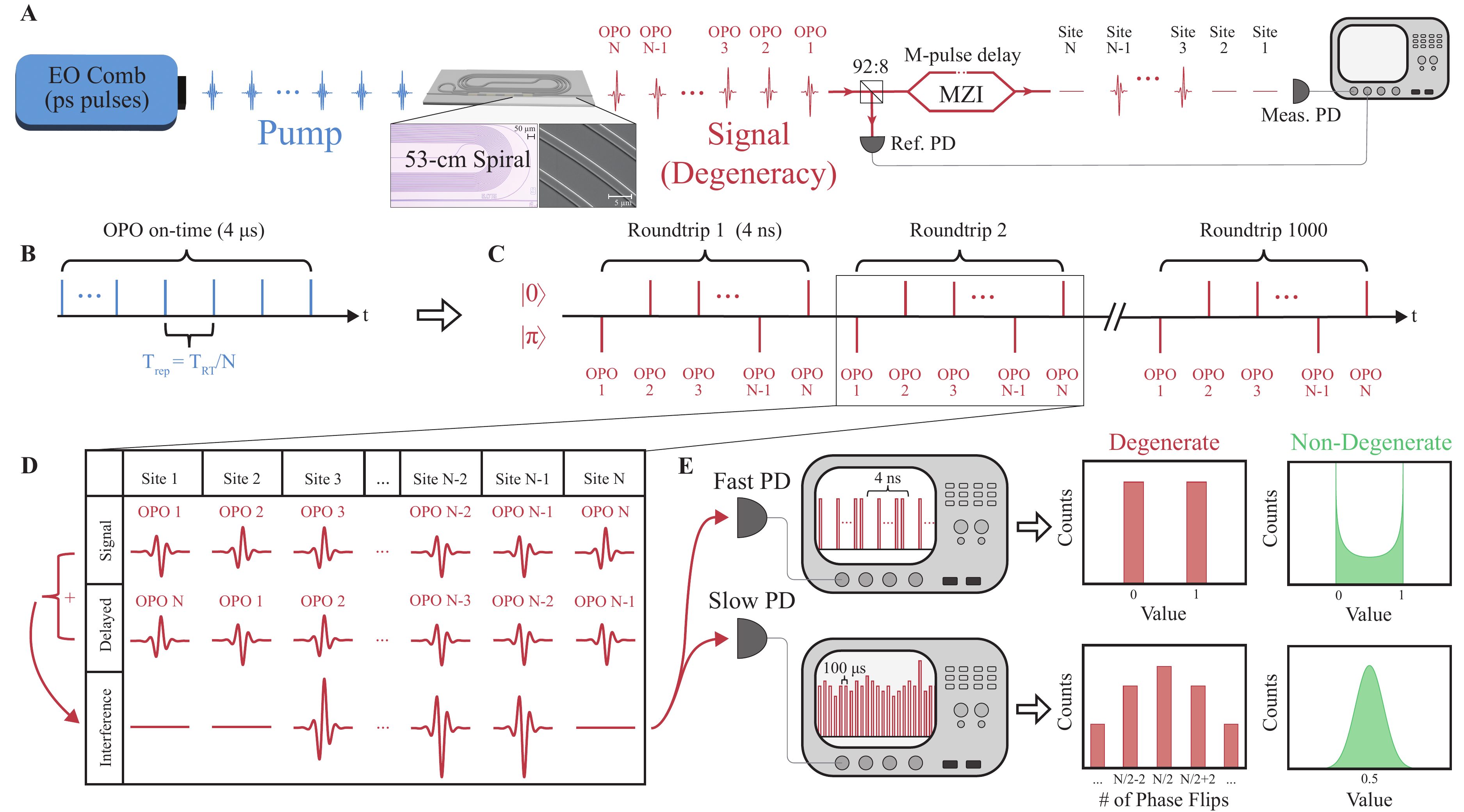}
\caption{{\bf Measurement of independent oscillators.} ({\bf A}) Schematic of the measurement setup. The nanophotonic chip is pumped by the output of an EO comb which provides ps pulses at GHz repetition rates. The output is passed through a fiber interferometer with an M-pulse delay for characterizing the relative phases of the output pulses. The inset shows microscope (left) and SEM (right) images of a segment of the spiral resonator. ({\bf B}) Illustration of the pump pulse train, with 4-\textmu s on-time. ({\bf C}) Example signal pulse train at degeneracy, showing the binary phase of the output. Each OPO iteration contains a repeating 4-ns random pattern of \(\ket{0}\) and \(\ket{\pi}\). ({\bf D}) Principle of the interferometric measurement of the OPO signal, illustrated with a 1-pulse delay. ({\bf E}) Expected outputs for pulse-to-pulse (fast) and average (slow) measurements. EO, electro-optic; PD, photodetector.}
\end{figure}

Figures 2B and 2C illustrate an example of the pump and signal pulse trains, respectively. The pump (Fig. 2B) consists of a train of ps pulses to pump N time-multiplexed OPOs which are kept on for 1000 roundtrips (4 \textmu s). The resulting signal (Fig. 2C) is composed of a train of N pulses with phases sampled randomly from the allowed phase states of the system. In the depicted case of degeneracy, they are sampled from the binary states \(\ket{0}\) and \(\ket{\pi}\).

Figure 2D depicts the passage of the signal through the MZI for each measurement. Here, a 1-pulse delay is assumed for simplicity. The top row shows the signal arm while the middle row shows the delayed arm. Their recombination results in the interference signal shown in the bottom row, which is finally sent to the measurement detector. In our experiment, this measurement is repeated every 100 \textmu s by turning the OPOs off and sending another 4-\textmu s set of pump pulses for collecting statistics. As shown in Fig. 2E, we have employed both a fast detector for shot-to-shot measurement of the resulting intensity in each pulse site and a slow detector for averaging over the MZI output for each 4-\textmu s set of pump pulses.

Each time the OPOs are turned on, each independent OPO signal takes a random phase. In the case of fast detector measurements (Fig. 2E, top) and degenerate operation, the corresponding normalized pulse peak intensity distribution taken across many OPO iterations should resemble a Bernoulli distribution with a 50\% probability for obtaining either 0 or 1, such that the probability mass function (PMF) is given by:

\begin{equation}
    f(x) = \begin{cases} 
        0.5 & x = 0, \\
        0.5 & x = 1, \\
        0 & \text{otherwise}.
    \end{cases}
\end{equation}

By contrast, in the non-degenerate case where the signal can take on an arbitrary phase, the expected theoretical distribution is given by the probability density function (PDF):

\begin{equation}
    f(x) = \begin{cases}
        \frac{1}  {\pi\sqrt{x(1-x)}} & 0 \leq x < 1, \\
        0 & \text{otherwise}.
        \end{cases}
\end{equation}

In the case of slow detector measurements (Fig. 2E, bottom), the average value of the MZI signal over the 4-ns pulse train should be considered. Such a measurement is helpful for obtaining statistics over a larger number of OPO iterations as well as for improving measurement SNR in the case where the OPO repetition rate approaches the detection bandwidth of the fast measurement system. In this case, the PMF for the degenerate case is expected to be:

\begin{equation}
    f(x) = \begin{cases}
        \frac{1}{2^{N-1}}{N \choose k} & k\ \text{even}, \\
        0 & \text{otherwise},
    \end{cases}
\end{equation}

\noindent where \(k\ \epsilon\ {0, 1, ... ,N}\) is the number of phase flips between consecutive pulses in the train of N pulses. Meanwhile, for large N, the non-degenerate system should tend towards a normal distribution. More information regarding the expected outputs can be found in Supplementary Section 2.

\begin{figure}[hp!]
\centering\includegraphics[width=16cm]{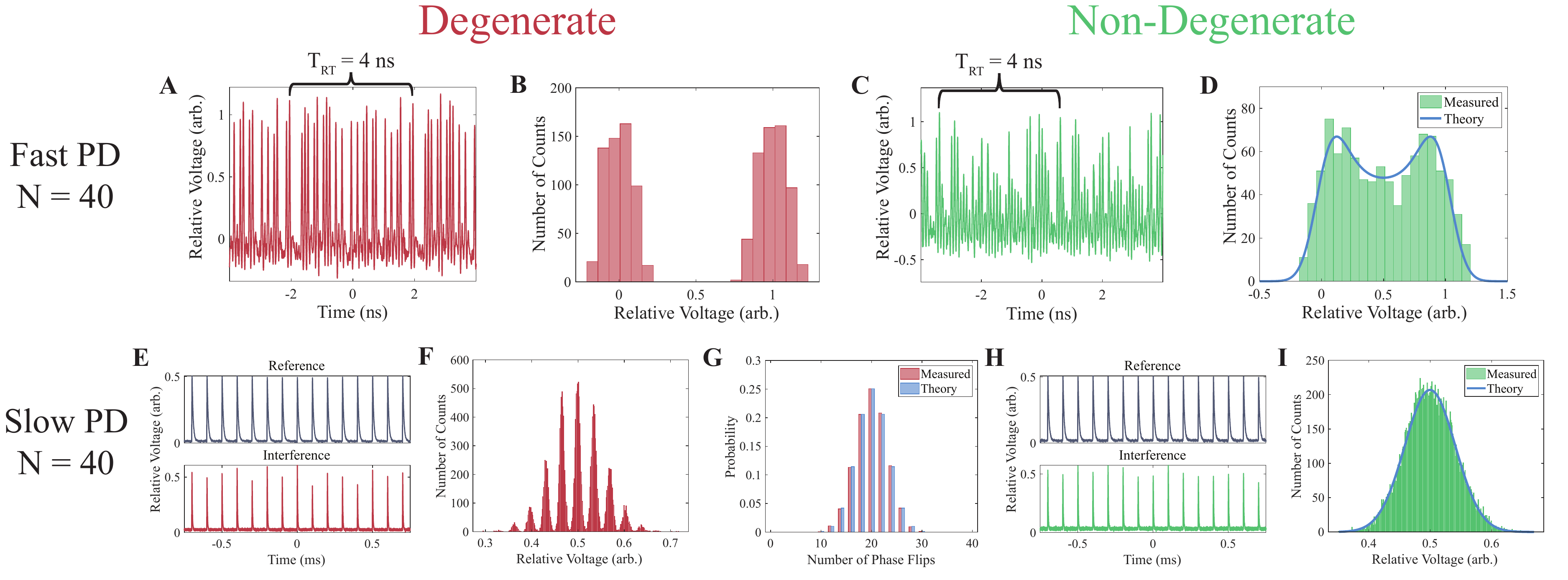}
\caption{{\bf Interference measurements for N = 40.} ({\bf A}-{\bf D}) Fast detector measurements in the degenerate (A-B) and non-degenerate (C-D) regimes. Examples of the measured interference pulse trains are shown in (A) and (C). Histograms of the measured peak pulse intensities over 30 degenerate (B) and 25 non-degenerate (D) interference measurements show good agreement with theoretical expectations. ({\bf E}-{\bf I}) Slow detector measurements in the degenerate (E-G) and non-degenerate (H-I) regimes. Example data out of the slow detector is shown in (E) and (H). The reference trace (top) is sampled from the OPO output before the MZI and is used for intensity noise correction of the interference trace (bottom). Histograms of the resulting corrected traces are shown in (F) and (I). As expected theoretically, discrete peaks are observed in the degenerate case (F), whereas a continuous distribution is seen in the the non-degenerate case (I). The discrete peaks in the degenerate case are binned for comparison with theory (G), showing good agreement.}
\end{figure}

We first measure the case where N = 40, meaning the repetition rate of the pump is set to be 10 GHz, the results of which are presented in Fig. 3. Figures 3A-D contain the results of the interferometric measurement on a 25-GHz fast detector, with Figs. 3A-B showing the output when the laser is detuned to degeneracy while Figs. 3C-D show the output in the non-degenerate case. Figures 3A and 3C show examples of the directly measured interference pulse trains. The y-axis shows the detector voltage, normalized such that 0 and 1 roughly correspond to the cases of constructive and destructive interference between the arms of the MZI, respectively. As expected theoretically, the MZI output is binary in the case of degeneracy (Fig. 3A). This is further confirmed by the histogram of the peak pulse intensities over 30 OPO iterations shown in Fig. 3B, where we observe two well-separated lobes around 1 and 0 with nearly equal probability. As discussed in Supplementary Section 2, one key difference between our measurement and theory is the addition of detector noise which results in the measured distribution looking like the convolution of the expected theoretical distribution with a Gaussian. Meanwhile, the MZI output in the non-degenerate regime can take on any value between 0 and 1 (Fig. 3C). By taking the histogram of the pulse peaks over 25 OPO iterations, we obtain the distribution shown in Fig. 3D. As expected, the measured distribution is bimodal, and it agrees well with the theoretical fit.

The independence of the pulses is further confirmed through the slow detector measurements shown in Fig. 3E-I. Figure 3E shows a snippet of the detected signal in the degenerate case. The reference measurement is acquired directly after the chip output and is used for intensity noise correction, while the interference data is taken after the MZI. Both are measured using a 1 MHz detector such that each of the observed peaks correspond to one OPO iteration. The histogram of Fig. 3F contains the peak values from the intensity-noise-corrected interference trace over 2 s of data, corresponding to 19,999 OPO iterations. Here, the observance of a discretization in the measured values confirms operation in the degenerate regime. Further comparison between the measurement and the expected theoretical distribution in Fig. 3G shows strikingly good agreement, verifying the independence of the OPOs in the time-multiplexed system.

The corresponding non-degenerate measurement is shown in Figs. 3H-I. Figure 3H shows a snippet of the data in this case. In contrast to the degenerate case, the interference output is not discretized, an observation confirmed by the histogram shown in Fig. 3I. Again, the histogram consists of 2 s of data. A Gaussian fit over the histogram shows good agreement between the measured data and the theoretically expected distribution.

\begin{figure}[hp!]
\centering\includegraphics[width=16cm]{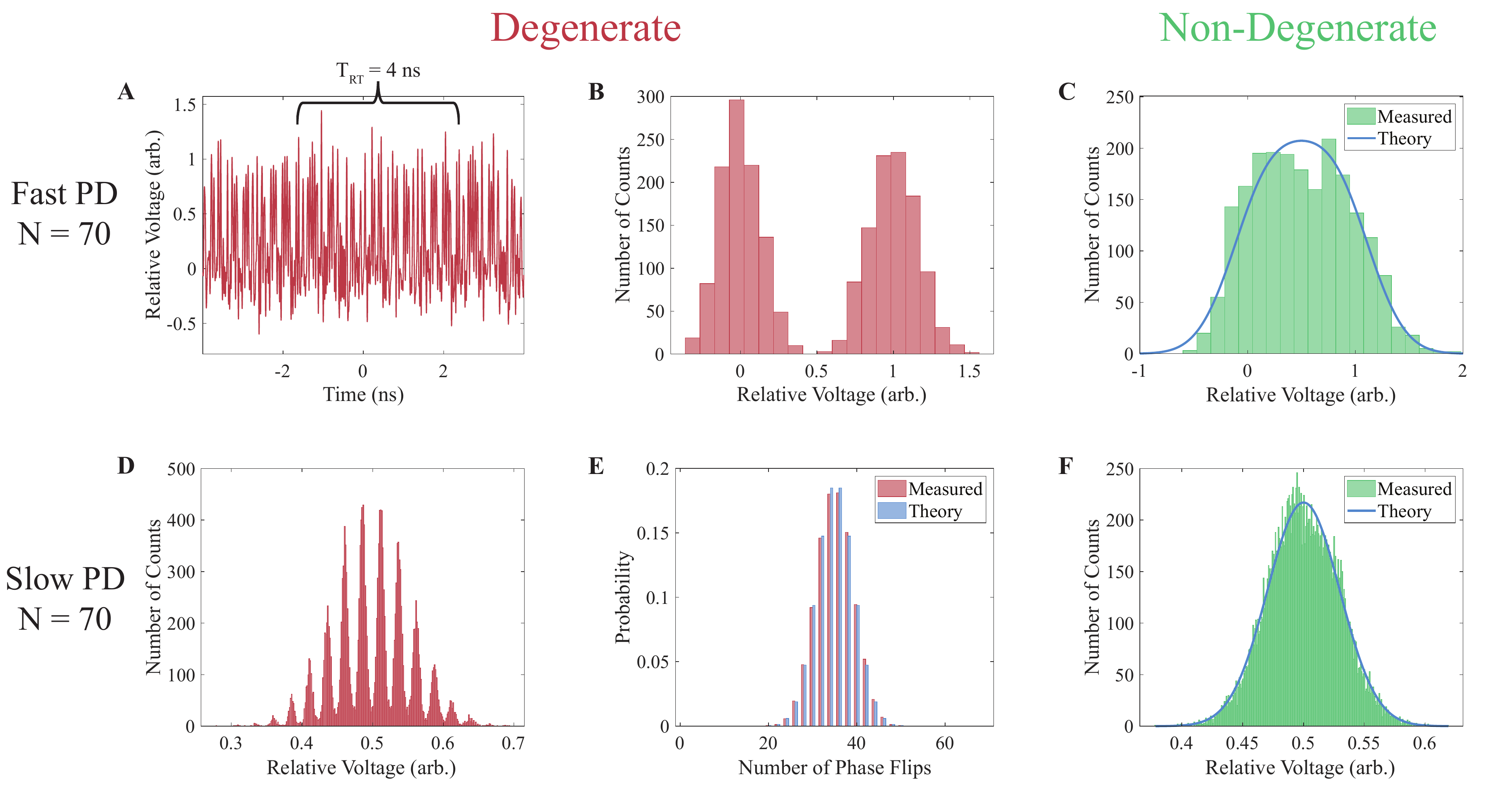}
\caption{{\bf Extension to N = 70.} ({\bf A}) Example inteference pulse train on the fast detector in the degenerate case. ({\bf B}) Histogram of peak pulse intensities over 30 degenerate OPO iterations, exhibiting the expected Bernoulli distribution. ({\bf C}) Corresponding histogram of fast detector measurements in the non-degenerate regime and theoretical fit. ({\bf D}) Histogram of slow detector measurements in the degenerate regime. ({\bf E}) Theoretical comparison after binning the distribution from (D), showing close agreement. ({\bf F}) Histogram of non-degenerate interference data and corresponding Gaussian fit.}
\end{figure}

One advantage of the TDM scheme is the ability to change the number of sites without significant additional overhead, making computation of problems of different sizes readily achievable. We demonstrate this by pushing our system to N = 70 at a pump repetition rate of 17.5 GHz. Our current measurement is limited by the modulators used in the EO comb and bandwidth of our fast photodetector. However, as faster nanophotonic pump sources become available, we believe the current time-multiplexed system will be capable of supporting much larger Ns and significantly faster pump rates.

Figure 4 presents the results for pumping with N = 70. As can be seen in the raw data of Fig. 4A measured with the fast detector, pumping near the limits of our electronics results in a reduced signal-to-noise compared to the case of N = 40. However, the histogram of fast detector measurements at degeneracy over 30 OPO iterations depicted in Fig. 4B again shows a clear Bernoulli distribution, in accordance with the theory. This stands in contrast with the non-degenerate case of Fig. 4C, which also agrees well with the theoretical fit. Here, the bimodality is less prevalent because of the relatively larger detector noise. The slow detector measurements in Figs. 4D-F further confirm that independence is maintained. At degeneracy, discrete values are again observed in the histogram (Fig. 4D), and the distribution is shown in Fig. 4E to match the theoretically expected distribution. Likewise, the non-degenerate measurement shown in Fig. 4F agrees well with the Gaussian fit. 

\section*{Outlook}

In this work, we have demonstrated a large-scale system of time-multiplexed OPOs in lithium niobate nanophotonics and shown the independence of the oscillators in both the degenerate and non-degenerate regimes through interferometric measurements of the device output. The addition of programmable couplings between the oscillators will allow for exploration of a multitude of fundamental phenomena and enable large-scale, all-optical information processing. Towards this end, we propose the architecture illustrated in Fig. 5A. Here, the main cavity consisting of N time-multiplexed OPOs is accompanied by a secondary memory cavity with N + 1 sites. The cavities are coupled such that pulses in the main and memory cavities can interact, and the strength of the coupling is set through MZI-based intensity modulators using fast electro-optic modulators (EOMs). A second pump and periodically poled region in the memory cavity is used to compensate the roundtrip loss.

\begin{figure}[hp!]
\centering\includegraphics[width=16cm]{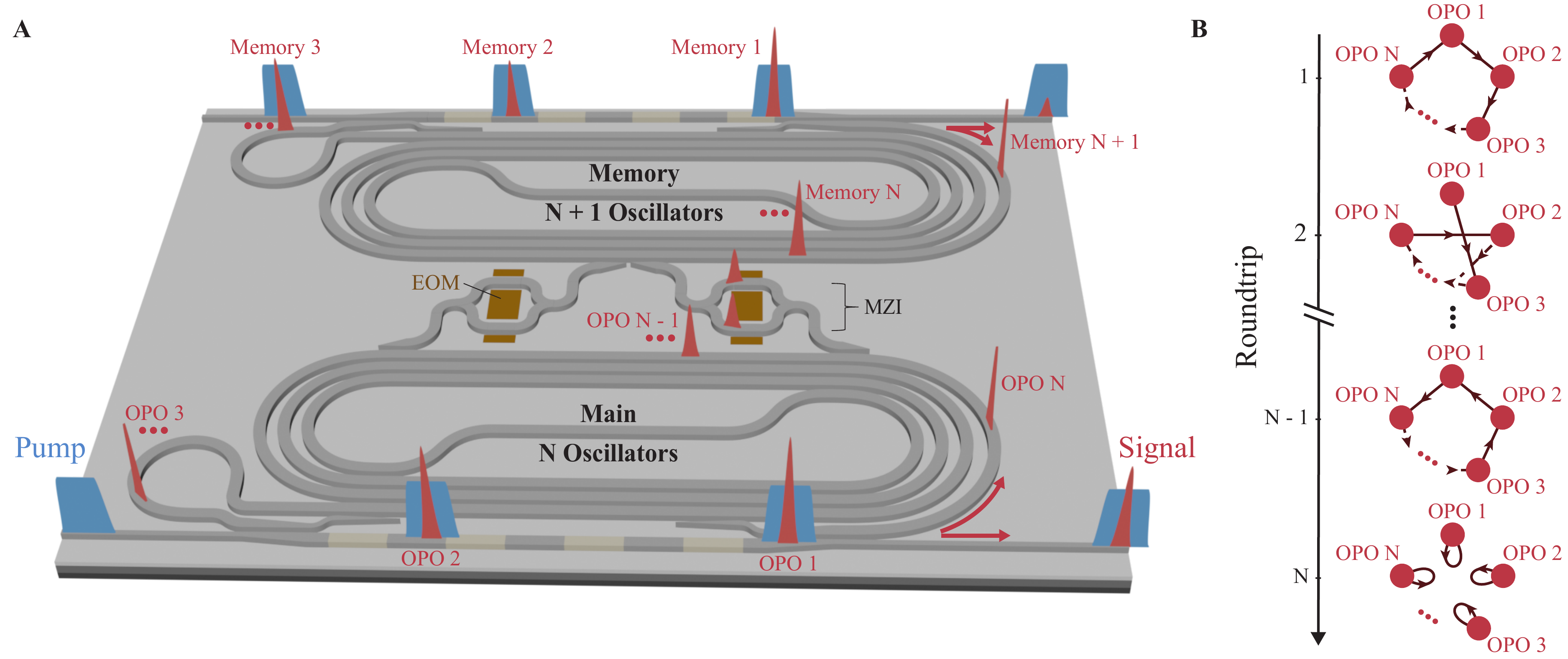}
\caption{{\bf Time-multiplexed architecture for programmable all-to-all coupled nonlinear resonators.} (\textbf{A}) The main resonator (bottom) with N time-multiplexed OPOs is coupled to a secondary memory cavity (top), designed to have N + 1 sites. Losses in the memory cavity are compensated by a secondary pump and poled region. EOMs in the coupling regions are used to program an MZI-based intensity modulator, which can be used to tune the strength of the coupled pulse. (\textbf{B}) Connectivity diagrams showing how arbitrary couplings are achieved over N roundtrips. EOM, electro-optic modulator; MZI, Mach-Zehnder interferometer.}
\end{figure}

In general, information processing requires a mixture of both linear and nonlinear operations. In this architecture, linear operations, such as multiply-accumulate operations or dot products, are achieved through arbitrary all-to-all couplings which may be implemented over N + 1 roundtrips of the memory cavity. In the first roundtrip, pulses from the main cavity are coupled to the memory cavity through the right MZI channel. In the subsequent N roundtrips, couplings occur as illustrated in the connectivity diagrams of Fig. 5B from the memory cavity back to the main cavity through the left MZI channel. The memory cavity may then be emptied of pulses to allow the next coupling cycle to begin. Meanwhile, all-optical nonlinear functions \cite{li2023all} may be applied by modulating the pump in the main cavity. Such a system, therefore, offers a flexible platform for studying systems of coupled nonlinear resonators and highlights the importance of our demonstrated time-multiplexed OPO as a first step towards achieving integrated ultrafast, energy-efficient, and scalable all-optical information processing systems.

\bibliographystyle{abbrv}
\bibliography{sample.bib}

\begin{thebibliography}{10}

\bibitem{arwas2022anyonic}
G.~Arwas, S.~Gadasi, I.~Gershenzon, A.~Friesem, N.~Davidson, and O.~Raz.
\newblock Anyonic-parity-time symmetry in complex-coupled lasers.
\newblock {\em Science advances}, 8(22):eabm7454, 2022.

\bibitem{asavanant2019generation}
W.~Asavanant, Y.~Shiozawa, S.~Yokoyama, B.~Charoensombutamon, H.~Emura, R.~N. Alexander, S.~Takeda, J.-i. Yoshikawa, N.~C. Menicucci, H.~Yonezawa, et~al.
\newblock Generation of time-domain-multiplexed two-dimensional cluster state.
\newblock {\em Science}, 366(6463):373--376, 2019.

\bibitem{bandres2018topological}
M.~A. Bandres, S.~Wittek, G.~Harari, M.~Parto, J.~Ren, M.~Segev, D.~N. Christodoulides, and M.~Khajavikhan.
\newblock Topological insulator laser: Experiments.
\newblock {\em Science}, 359(6381):eaar4005, 2018.

\bibitem{dunn1999parametric}
M.~H. Dunn and M.~Ebrahimzadeh.
\newblock Parametric generation of tunable light from continuous-wave to femtosecond pulses.
\newblock {\em Science}, 286(5444):1513--1517, 1999.

\bibitem{gray2023cavity}
R.~M. Gray, M.~Liu, S.~Zhou, A.~Roy, L.~Ledezma, and A.~Marandi.
\newblock Cavity-soliton-enhanced mid-ir molecular sensing.
\newblock {\em arXiv preprint arXiv:2301.07826}, 2023.

\bibitem{honjo2021100}
T.~Honjo, T.~Sonobe, K.~Inaba, T.~Inagaki, T.~Ikuta, Y.~Yamada, T.~Kazama, K.~Enbutsu, T.~Umeki, R.~Kasahara, et~al.
\newblock 100,000-spin coherent ising machine.
\newblock {\em Science advances}, 7(40):eabh0952, 2021.

\bibitem{jang2018synchronization}
J.~K. Jang, A.~Klenner, X.~Ji, Y.~Okawachi, M.~Lipson, and A.~L. Gaeta.
\newblock Synchronization of coupled optical microresonators.
\newblock {\em Nature Photonics}, 12(11):688--693, 2018.

\bibitem{konno2024logical}
S.~Konno, W.~Asavanant, F.~Hanamura, H.~Nagayoshi, K.~Fukui, A.~Sakaguchi, R.~Ide, F.~China, M.~Yabuno, S.~Miki, et~al.
\newblock Logical states for fault-tolerant quantum computation with propagating light.
\newblock {\em Science}, 383(6680):289--293, 2024.

\bibitem{larsen2019deterministic}
M.~V. Larsen, X.~Guo, C.~R. Breum, J.~S. Neergaard-Nielsen, and U.~L. Andersen.
\newblock Deterministic generation of a two-dimensional cluster state.
\newblock {\em Science}, 366(6463):369--372, 2019.

\bibitem{ledezma2022intense}
L.~Ledezma, R.~Sekine, Q.~Guo, R.~Nehra, S.~Jahani, and A.~Marandi.
\newblock Intense optical parametric amplification in dispersion-engineered nanophotonic lithium niobate waveguides.
\newblock {\em Optica}, 9(3):303--308, 2022.

\bibitem{leefmans2022topological}
C.~Leefmans, A.~Dutt, J.~Williams, L.~Yuan, M.~Parto, F.~Nori, S.~Fan, and A.~Marandi.
\newblock Topological dissipation in a time-multiplexed photonic resonator network.
\newblock {\em Nature Physics}, 18(4):442--449, 2022.

\bibitem{leefmans2022topologicalmode}
C.~Leefmans, M.~Parto, J.~Williams, G.~H. Li, A.~Dutt, F.~Nori, and A.~Marandi.
\newblock Topological temporally mode-locked laser.
\newblock {\em arXiv preprint arXiv:2209.00762}, 2022.

\bibitem{li2023photonic}
G.~H. Li, C.~R. Leefmans, J.~Williams, and A.~Marandi.
\newblock Photonic elementary cellular automata for simulation of complex phenomena.
\newblock {\em Light: Science \& Applications}, 12(1):132, 2023.

\bibitem{li2023all}
G.~H. Li, R.~Sekine, R.~Nehra, R.~M. Gray, L.~Ledezma, Q.~Guo, and A.~Marandi.
\newblock All-optical ultrafast relu function for energy-efficient nanophotonic deep learning.
\newblock {\em Nanophotonics}, 12(5):847--855, 2023.

\bibitem{lu2021ultralow}
J.~Lu, A.~Al~Sayem, Z.~Gong, J.~B. Surya, C.-L. Zou, and H.~X. Tang.
\newblock Ultralow-threshold thin-film lithium niobate optical parametric oscillator.
\newblock {\em Optica}, 8(4):539--544, 2021.

\bibitem{marandi2012all}
A.~Marandi, N.~C. Leindecker, K.~L. Vodopyanov, and R.~L. Byer.
\newblock All-optical quantum random bit generation from intrinsically binary phase of parametric oscillators.
\newblock {\em Optics express}, 20(17):19322--19330, 2012.

\bibitem{marandi2014network}
A.~Marandi, Z.~Wang, K.~Takata, R.~L. Byer, and Y.~Yamamoto.
\newblock Network of time-multiplexed optical parametric oscillators as a coherent ising machine.
\newblock {\em Nature Photonics}, 8(12):937--942, 2014.

\bibitem{mckenna2022ultra}
T.~P. McKenna, H.~S. Stokowski, V.~Ansari, J.~Mishra, M.~Jankowski, C.~J. Sarabalis, J.~F. Herrmann, C.~Langrock, M.~M. Fejer, and A.~H. Safavi-Naeini.
\newblock Ultra-low-power second-order nonlinear optics on a chip.
\newblock {\em Nature Communications}, 13(1):4532, 2022.

\bibitem{mcmahon2016fully}
P.~L. McMahon, A.~Marandi, Y.~Haribara, R.~Hamerly, C.~Langrock, S.~Tamate, T.~Inagaki, H.~Takesue, S.~Utsunomiya, K.~Aihara, et~al.
\newblock A fully programmable 100-spin coherent ising machine with all-to-all connections.
\newblock {\em Science}, 354(6312):614--617, 2016.

\bibitem{mittal2021topological}
S.~Mittal, G.~Moille, K.~Srinivasan, Y.~K. Chembo, and M.~Hafezi.
\newblock Topological frequency combs and nested temporal solitons.
\newblock {\em Nature Physics}, 17(10):1169--1176, 2021.

\bibitem{muraviev2018massively}
A.~Muraviev, V.~O. Smolski, Z.~E. Loparo, and K.~L. Vodopyanov.
\newblock Massively parallel sensing of trace molecules and their isotopologues with broadband subharmonic mid-infrared frequency combs.
\newblock {\em Nature Photonics}, 12(4):209--214, 2018.

\bibitem{obrzud2019microphotonic}
E.~Obrzud, M.~Rainer, A.~Harutyunyan, M.~H. Anderson, J.~Liu, M.~Geiselmann, B.~Chazelas, S.~Kundermann, S.~Lecomte, M.~Cecconi, et~al.
\newblock A microphotonic astrocomb.
\newblock {\em Nature Photonics}, 13(1):31--35, 2019.

\bibitem{okawachi2020demonstration}
Y.~Okawachi, M.~Yu, J.~K. Jang, X.~Ji, Y.~Zhao, B.~Y. Kim, M.~Lipson, and A.~L. Gaeta.
\newblock Demonstration of chip-based coupled degenerate optical parametric oscillators for realizing a nanophotonic spin-glass.
\newblock {\em Nature communications}, 11(1):4119, 2020.

\bibitem{ozawa2019topological}
T.~Ozawa and H.~M. Price.
\newblock Topological quantum matter in synthetic dimensions.
\newblock {\em Nature Reviews Physics}, 1(5):349--357, 2019.

\bibitem{pal2024linear}
A.~Pal, A.~Ghosh, S.~Zhang, L.~Hill, H.~Yan, H.~Zhang, T.~Bi, A.~Alabbadi, and P.~Del'Haye.
\newblock Linear and nonlinear coupling of twin-resonators with kerr nonlinearity.
\newblock {\em arXiv preprint arXiv:2404.05646}, 2024.

\bibitem{parto2023non}
M.~Parto, C.~Leefmans, J.~Williams, F.~Nori, and A.~Marandi.
\newblock Non-abelian effects in dissipative photonic topological lattices.
\newblock {\em Nature Communications}, 14(1):1440, 2023.

\bibitem{roques2023biasing}
C.~Roques-Carmes, Y.~Salamin, J.~Sloan, S.~Choi, G.~Velez, E.~Koskas, N.~Rivera, S.~E. Kooi, J.~D. Joannopoulos, and M.~Solja{\v{c}}i{\'c}.
\newblock Biasing the quantum vacuum to control macroscopic probability distributions.
\newblock {\em Science}, 381(6654):205--209, 2023.

\bibitem{roy2023visible}
A.~Roy, L.~Ledezma, L.~Costa, R.~Gray, R.~Sekine, Q.~Guo, M.~Liu, R.~M. Briggs, and A.~Marandi.
\newblock Visible-to-mid-ir tunable frequency comb in nanophotonics.
\newblock {\em Nature Communications}, 14(1):6549, 2023.

\bibitem{roy2023non}
A.~Roy, R.~Nehra, C.~Langrock, M.~Fejer, and A.~Marandi.
\newblock Non-equilibrium spectral phase transitions in coupled nonlinear optical resonators.
\newblock {\em Nature Physics}, 19(3):427--434, 2023.

\bibitem{sekine2023multi}
R.~Sekine, R.~M. Gray, L.~Ledezma, S.~Zhou, Q.~Guo, and A.~Marandi.
\newblock Multi-octave frequency comb from an ultra-low-threshold nanophotonic parametric oscillator.
\newblock {\em arXiv preprint arXiv:2309.04545}, 2023.

\bibitem{spencer2018optical}
D.~T. Spencer, T.~Drake, T.~C. Briles, J.~Stone, L.~C. Sinclair, C.~Fredrick, Q.~Li, D.~Westly, B.~R. Ilic, A.~Bluestone, et~al.
\newblock An optical-frequency synthesizer using integrated photonics.
\newblock {\em Nature}, 557(7703):81--85, 2018.

\bibitem{stern2020direct}
L.~Stern, J.~R. Stone, S.~Kang, D.~C. Cole, M.-G. Suh, C.~Fredrick, Z.~Newman, K.~Vahala, J.~Kitching, S.~A. Diddams, et~al.
\newblock Direct kerr frequency comb atomic spectroscopy and stabilization.
\newblock {\em Science advances}, 6(9):eaax6230, 2020.

\bibitem{takeda2017boltzmann}
Y.~Takeda, S.~Tamate, Y.~Yamamoto, H.~Takesue, T.~Inagaki, and S.~Utsunomiya.
\newblock Boltzmann sampling for an xy model using a non-degenerate optical parametric oscillator network.
\newblock {\em Quantum Science and Technology}, 3(1):014004, 2017.

\bibitem{wang2013coherent}
Z.~Wang, A.~Marandi, K.~Wen, R.~L. Byer, and Y.~Yamamoto.
\newblock Coherent ising machine based on degenerate optical parametric oscillators.
\newblock {\em Physical Review A}, 88(6):063853, 2013.

\bibitem{yamamoto2017coherent}
Y.~Yamamoto, K.~Aihara, T.~Leleu, K.-i. Kawarabayashi, S.~Kako, M.~Fejer, K.~Inoue, and H.~Takesue.
\newblock Coherent ising machines—optical neural networks operating at the quantum limit.
\newblock {\em npj Quantum Information}, 3(1):49, 2017.

\bibitem{yanagimoto2023mesoscopic}
R.~Yanagimoto, E.~Ng, M.~Jankowski, R.~Nehra, T.~P. McKenna, T.~Onodera, L.~G. Wright, R.~Hamerly, A.~Marandi, M.~Fejer, et~al.
\newblock Mesoscopic ultrafast nonlinear optics--the emergence of multimode quantum non-gaussian physics.
\newblock {\em arXiv preprint arXiv:2311.13775}, 2023.

\bibitem{zhang2019electronically}
M.~Zhang, C.~Wang, Y.~Hu, A.~Shams-Ansari, T.~Ren, S.~Fan, and M.~Lon{\v{c}}ar.
\newblock Electronically programmable photonic molecule.
\newblock {\em Nature Photonics}, 13(1):36--40, 2019.

\bibitem{zhang2013mid}
Z.~Zhang, T.~Gardiner, and D.~T. Reid.
\newblock Mid-infrared dual-comb spectroscopy with an optical parametric oscillator.
\newblock {\em Optics letters}, 38(16):3148--3150, 2013.

\end{thebibliography}

\section*{Acknowledgments}

The device nanofabrication was performed at the Kavli Nanoscience Institute (KNI) at Caltech. The authors acknowledge support from ARO grant no. W911NF-23-1-0048, NSF grant no. 1846273 and 1918549, AFOSR award FA9550-23-1-0755, DARPA award D23AP00158, Center for Sensing to Intelligence at Caltech, and NTT Research. G.H.Y.L. acknowledges support from the Quad Fellowship. R.M.G. gratefully acknowledges support from the NSF Graduate Research Fellowship Program.

\paragraph{Author Contributions:}
R.M.G. and A.M. conceived the idea and designed the experiments. R.M.G. and R.S. performed the experiments with help from L.L and A.R. R.S. fabricated the device with help from S.Z. R.M.G. performed the theoretical analysis and data processing. R.S., G.H.Y.L., M.P., and A.M. developed the all-to-all coupled architecture. R.M.G. and A.M. wrote the manuscript with input from all authors. A.M. supervised the project.

\paragraph{Competing Interests:}
L.L. and A.M. are inventors on granted U.S. patent 11,226,538. R.S., R.M.G., L.L., A.R., and A.M. are inventors on a U.S. provisional patent application filed by the California Institute of Technology (application number 63/466,188). G.H.Y.L., M.P., R.S., and A.M. are inventors on the US patent application number 18/448700. L.L. and A.M. are involved in developing photonic integrated nonlinear circuits at PINC Technologies Inc. L.L. and A.M. have an equity interest in PINC Technologies Inc.

\paragraph{Data Availability:}
The data used for generation of the figures within this manuscript and other findings of this study are available upon request from the corresponding author.

\paragraph{Code Availability:}
The code used for simulation and plotting of results is available upon request from the corresponding author.

\end{document}